\documentclass[conference,10pt,letterpaper]{IEEEtran}
\IEEEoverridecommandlockouts
\usepackage{cite}
\usepackage{amsmath,amssymb,amsfonts}
\usepackage{algorithmic}
\usepackage{graphicx}
\usepackage{subcaption}
\usepackage{textcomp}
\usepackage{xcolor}
\usepackage{todonotes}
\usepackage{array}
\usepackage{booktabs}
\def\BibTeX{{\rm B\kern-.05em{\sc i\kern-.025em b}\kern-.08em
    T\kern-.1667em\lower.7ex\hbox{E}\kern-.125emX}}

\newcommand{\sspace}[0]{\vspace{14pt}}

\usepackage{hyperref}

\begin{document}

\title{\fontsize{24}{24}\selectfont{A System Level Performance Evaluation for Superconducting Digital Systems}}
\author{\fontsize{11}{11}\selectfont Joyjit Kundu, Debjyoti Bhattacharjee, Nathan Josephsen, Ankit Pokhrel, Udara De Silva, Wenzhe Guo, \\ Steven Van Winckel, Steven Brebels, Manu Perumkunnil, Quentin Herr and Anna Herr\\
\fontsize{10}{12}\selectfont IMEC}

\maketitle
\noindent \textbf{\textit{Abstract}}-- Superconducting Digital~(SCD) technology offers significant potential for enhancing the performance of next generation large scale compute workloads. By leveraging advanced lithography and a 300 mm platform, SCD devices can reduce energy consumption and boost computational power. This paper presents a cross-layer modeling approach to evaluate the system-level performance benefits of SCD architectures for Large Language Model~(LLM) training and inference. Our findings, based on experimental data and Pulse Conserving Logic~(PCL) design principles, demonstrate substantial performance gain in both training and inference. We are, thus, able to convincingly show that the SCD technology can address memory and interconnect limitations of present day solutions for next-generation compute systems.
\vspace{-0.5ex}


\section{Introduction}
\noindent The ever increasing demand for Generative Artificial Intelligence is shaping the roadmap of next generation compute systems~\cite{McKinsey_LLM,HPCwire_LLM}. Large language models~(LLMs) in particular have dominated this domain given their versatility and  performance scaling~\cite{transformer_paper}. However, training typical LLMs is very compute intensive and requires processing a huge amount of data with significant compute time resulting in an unprecedented level of carbon footprint~\cite{NIST_Energy}. For instance, training a GPT-3 model costs around 10M~USD with an estimated energy consumption of \(\sim\)1300~MWh\cite{paterson2022, paterson2021}. While LLM inference is computationally cheaper compared to training, the cumulative cost is extremely high in the long run, as multiple users queries are serviced~\cite{Inference_cost}. This is obviously not sustainable given the trend of increasing LLM size~\cite{AI_Memory}. Furthermore, the astronomical total cost of ownership~(TCO) for advanced CMOS technology~\cite{chip_TCO} used to build training and inference hardware, only serves to exacerbate the problem. 
Finally, semiconductor technology constraints like memory and interconnect scaling walls~\cite{Memory_wall,Interconnect_wall} adversely affect future AI models that will become increasingly memory or communication bound~\cite{AI_Memory,Scaling_laws}, requiring massive bandwidths to feed compute. Thus, there is a desperate need for a paradigm shift in compute system technology to tackle this global challenge.

Superconducting Digital (SCD) Electronics~~\cite{29_pcl_imec_asc_22,27_sc_processor_2018,30_scinterposer_2022,bairamkulov2024superconductive} has been one of the promising beyond-CMOS technologies with a potential for tremendous improvements across the whole stack. Unique physics and material advantages facilitate active devices with `sub-attoJoule' energy scales (at `ps' time scales), quantum accurate encoding of digital information~\cite{29_pcl_imec_asc_22,27_sc_processor_2018,30_scinterposer_2022}, wires with negligible dissipation and dispersion up to 100’s~GHz frequencies. The energy dissipation per switching event in SCD does not depend on the process node (like CMOS) but instead is relative to thermal noise.
Given these benefits, superconducting architectures stand out by being able to operate at \(\sim\)$20\times$ higher frequencies for a fraction of the on-chip power ($100\times$ less)~\cite{ayala2020mana,herr2023superconducting} and have $10000\times$ more energy efficient communication at the on-chip clock rate~\cite{egan2022synchronous,dai2022isochronous}. Cryogenic cooling can be applied to the entire system as opposed to individual dies, resulting in extremely high volumetric packaging density with all components of a system being both physically and electrically close~\cite{herr2024data}.

Despite such promising device data, it is unclear how such technology primitives would pan out when estimating overall performance while designing a full system for future AI or HPC workloads. 
It is imperative to perform early cross-layer system level evaluations to understand the competitiveness of SCD technology, compared to existing mature technologies. Our key contributions here are as follows: 
\begin{itemize}
    \item Construction of SCD-based system-architecture for LLM training and inference, via parametric architectural building blocks in a bottom-up manner, from SCD devices, pulse-conserving logic~(PCL), superconducting Josephson SRAM (JSRAM) and packaging solutions.
    \item Performance projection of the SCD system architecture for LLM training and inference based on an analytical performance modeling framework. 
    \item Showcasing the potential improvements when using SCD system compared to contemporary \mbox{GPU systems}.
\end{itemize}  

\begin{figure*}[t]
    \centering
\includegraphics[width=0.98\linewidth]{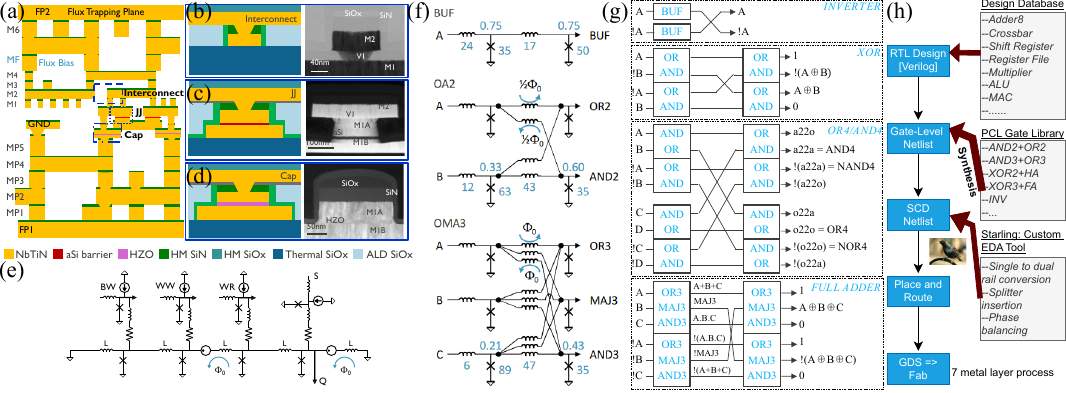}
    \caption{(a)Schematic of target 16ML SCD stack. (b)Scheme of 2ML BEOL interconnects and its TEM image. (c)Scheme of JJs with \(\alpha\)Si barriers and its TEM image. (d)Scheme of HZO MIM capacitor its TEM image. (e)HD JSRAM 1R/1W unit cell with 8 JJs (f)Building blocks of the PCL logic family. (g)Dual rail logic gates in the PCL cell library. (h)Outline of RTL-GDS automated flow.}
    \label{fig:tech_stack}
\end{figure*}

\section{\textbf{Background}}\label{sec:background}
\subsection{Advanced SCD process technology}
\noindent The fabrication of \texttt{Nb} based SCD devices have been reported previously. However, these have a temperature budget of \(\leq\)200\(^{\circ}\)C which limits effective scaling to lower dimensions for advanced process nodes, thus not meeting the process specifications required for complex integrated circuits~\cite{Nb_JJ1, Nb_JJ2, Nb_JJ3}. To address these challenges, we have developed \texttt{NbTiN} based SCD building blocks that allow advanced integration (meeting the requirements for advanced digital systems). Compared to the SOA \texttt{Nb}, \texttt{NbTiN} is a stable material, compatible with temperature budget of 420\(^{\circ}\)C and is scalable down to critical dimensions (CDs) of \(<\)50nm \cite{JJscaling1}\cite{JJscaling2}. Specifically, we use a legacy semi-damascene \cite{SCDprocess}\cite{SCDprocess2} integration process and report the fabrication of \texttt{NbTiN} based BEOL interconnects, \texttt{NbTiN/\(\alpha\)Si/NbTiN} Josephson junctions~(JJs), and \texttt{NbTiN/HZO/NbTiN} tunable Metal-Insulator-Metal~(MIM) capacitors that are the fundamental blocks of SCD technology. The stack utilizes a standard interlayer dielectric process and 193i lithography (suitable for 40nm and 28nm) to achieve a device density of 400M JJs/cm$^2$. Measurements have demonstrated the scaling of BEOL interconnects to CDs of 50-600~nm, Amorphous \texttt{Si (\(\alpha\)Si)} based JJs to diameters between 210-500nm (CD control of \(\sigma\)\(<2\%\) across 300mm wafer), and tunable HZO MIM capacitors to diameters between 195-600nm (CD control of \(\sigma\)\(<2\%\) across 300 mm wafer). The resonant
AC network for power distribution is enabled by NbTiN wires and HZO MIM capacitors~\cite{herr2023scaling}. The process and device specifications are detailed in Fig.~\ref{fig:tech_stack} and Table.~\ref{table:process_spec}. 


\subsection{Pulse Conserving Logic and JSRAM}
\noindent Digital circuits have been constructed in various low-power superconducting logic families like RSFQ, RQL, and AQFP~\cite{bairamkulov2024superconductive}. These efforts have been held back by the lack of large scale design automation targeting superconducting logic. Superconducting circuits that are DC-powered waste energy in power distribution, while circuits that are AC-powered face challenges in static timing analysis and automated synthesis.

Pulse-conserving logic~(PCL)~\cite{herr2023superconducting} is a SCD logic family that enables automated circuit design using commercially available tools. In a PCL circuit, each digital signal comprises two physical wires (positive and negative sense) and inversion is achieved by swapping them. This dual-rail encoding eliminates the delay in logical inversion (inherent to the data encoding in other AC powered SCD logic families) and greatly simplifies mapping of PCL gates into a standard cell library, a subset of which is shown in Fig.~\ref{fig:tech_stack}f and \ref{fig:tech_stack}g. The PCL library is compatible with a slightly modified RTL-GDS flow. Fig.~\ref{fig:tech_stack}h illustrates the translation of an RTL design to a PCL circuit using off-the-shelf synthesis followed by a modification stage that performs phase assignment, phase matching, and single-to-dual rail conversion. A commercial place and route solution that can route wires with targeted inductance was used. JSRAM~\cite{herr2023superconducting} is a SCD memory technology complementary to PCL, which, implemented in the advanced \texttt{NbTiN/\(\alpha\)Si/NbTiN} process described in the previous subsection, enables memory density of $4$~MB/cm$^2$, a $600\times$ increase over older SFQ-compatible memory technology, with XY addressing analogous to CMOS SRAM. It represents a key intermediate stage of the memory hierarchy with lower density but higher locality than cryo-DRAM. 
\begin{table}[t]
\vspace{0.8cm}
 \centering
 \caption{Specifications for the SCD technology stack in our work}
 \label{table:process_spec}
 {\footnotesize
   \begin{tabular}{l|r|r}
  \hline
  \textbf{Parameter} & \textbf{CMOS 5nm} & \textbf{This work} \\\hline
  \textbf{Operating Frequency} & 2GHz & 30GHz \\ \hline
  \textbf{Device} & FinFET & Josephson Junction \\
  -- \textbf{Device Density} & \(\sim 170M/mm^2\) & \(\sim 4M/mm^2\) \\ 
  -- \textbf{Voltage} & \(0.7V\) & \(\sim\) 1.0mV \\ \hline
  \textbf{On-chip Memory} & SRAM & JSRAM   \\ \hline
  -- \textbf{Density}  (incl. peri) & \(\sim\)\(4.5MB/mm^2\) & \(\sim\)\(0.4M/mm^2\) \\ 
  -- \textbf{HD Device Unit Cell} & 6T & 8JJ \\ 
 (Single Port 1R/1W) & \(0.021\mu\)\(m^2\) & \(1.86\mu\)\(m^2\) \\ \hline
  \textbf{Lithography} & EUV & 193 \\ \hline
  \textbf{ML stack layers} & 16 & 16 \\ \hline
  \textbf{Interconnects} & Cu & NbTiN \\ 
  -- \textbf{Resisitivity} (M1-M3) & \(\sim\)75\(\mu\)\(\Omega\).m & \(<\)2\(\mu\)\(\Omega\).m \\ 
  -- \textbf{Minimum MP} & 28nm/35nm & 50nm \\ 
  -- \textbf{Power Efficiency} & 1-2Gb@1pJ/bit & \(\sim\)200Gb@1pJ/bit \\ \hline
 \end{tabular}
 }
 \vspace{-0.3cm}
 \end{table}

\section{\textbf{Design Specifications}}\label{sec:designspec}
\noindent In this section, we describe the fundamental design blocks that serve as inputs towards building the SCD blade for LLM training and inference.

\underline{\textbf{High Throughput Compute Core}}: A regular array of bf16 MAC units is used for a TPU like high throughput compute core. Our bf16 MAC (8-bit add, 8-bit multiply and 32 bit accumulate) consists of \(\sim\)8k JJs. This regular MAC array architecture is banked for scale-out. Due to the lack of RC overhead for superconducting data transmission lines and utilization of dedicated low latency memory hierarchy (detailed below), we achieve a very high packing density and utilization for the high throughput compute core. The peak floating point~(bf16) performance achieved is \(\sim\)2.45~PetaFLOPs~(30GHz$\times$400k MACs$\times$2Ops/MAC) at 80\(\%\) utilization of the MACs (\(>\)400K) in a 144\(mm^2\) die~footprint.

\underline{\textbf{Control Complex}}: A simple dual core (in-order) complex manages the distribution of kernels fragments and appropriate instructions to the high throughput core. The `Control Complex' maintains local directories for coherency for the global addressing. It also assists in power/clock gating (if any) locally.

\underline{\textbf{Memory Hierarchy}}: The 8JJ single port (1R/1W) JSRAM configuration detailed in Fig.~\ref{fig:tech_stack}e serves as the `High Density' (\textbf{HD}) on-chip memory variant for our SCD stack. HD JSRAM is used for the L1 and L2 data caches in our study. Multi-port variants, 2R/1W JSRAM (14 JJs) and 3R/2W (29 JJs), serve as the `High Performance' (\textbf{HP}) on-chip memory variants. These are used for register files, high speed buffers and L1 instruction caches in our system configurations based on the required parallelism, performance and capacity specifications.

\underline{\textbf{Switch}}: Like conventional designs, our SCD switch consists of a central crossbar that connect the input ports (\(+\)~associated buffers) to the control unit and output ports (\(+\) associated buffers). The building block of the crossbar is in-turn the superconducting MUX based cross-point unit. Our crossbar is hierarchical in nature with a first level of cross-point units used to route each packet to the appropriate output port, and the second level serving as an aggregation point.

\begin{figure}[t]
\centering
\begin{subfigure}[t]{0.38\linewidth}
    \caption{}
    \label{fig:datalink}
    \sspace
\includegraphics[width=\linewidth]{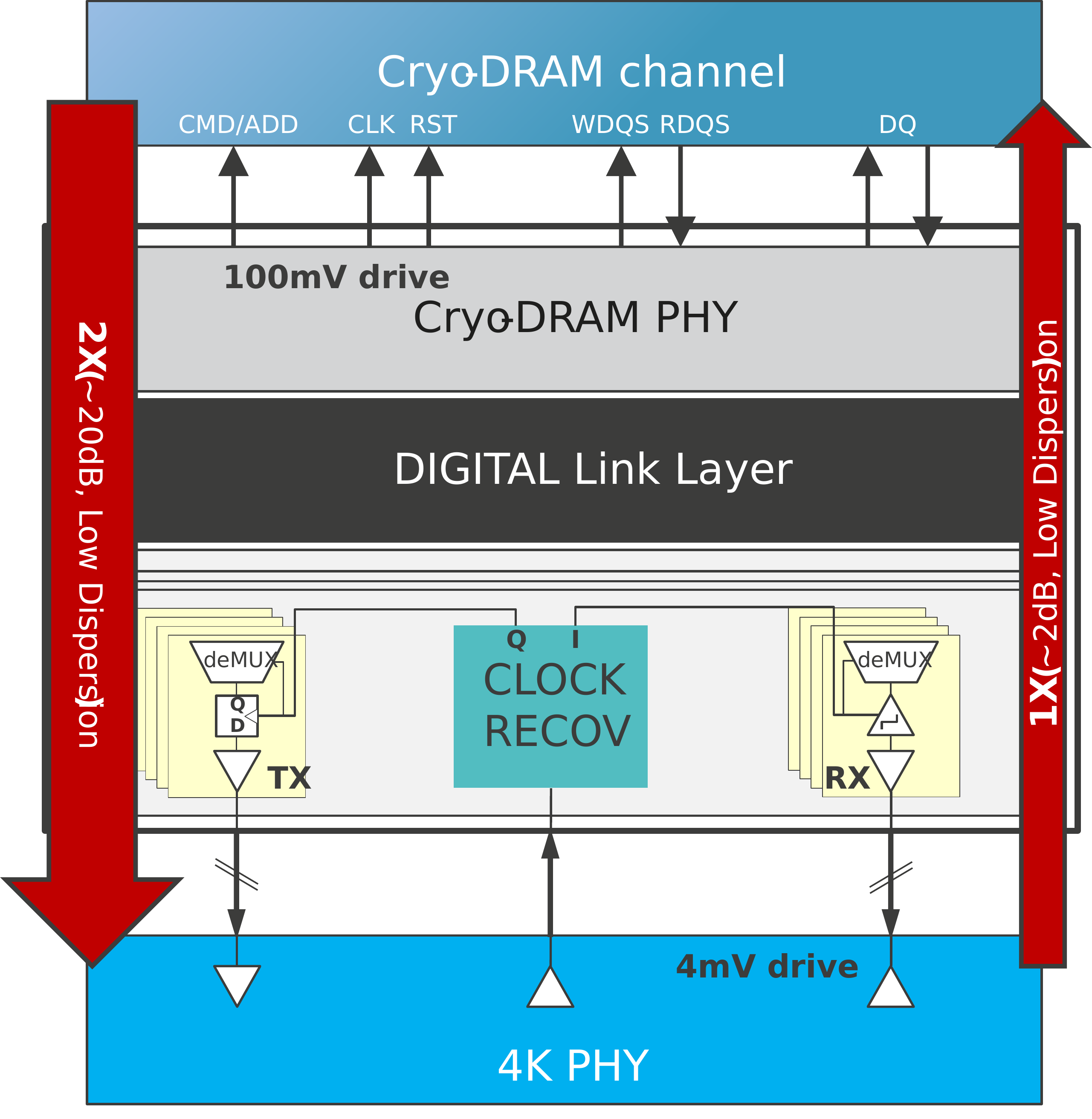}
\end{subfigure}
\begin{subfigure}[t]{0.6\linewidth}
\caption{}
\label{table:datalink_spec}
\sspace
\scalebox{0.85}{\footnotesize
  \begin{tabular}{l|r|r}
 \bottomrule
 \textbf{Parameter} & \textbf{Downlink} & \textbf{Uplink} \\
 & towards 4K & towards 77K \\ \midrule\midrule
 \textbf{Wire Width} & 6.2\(\mu\)m & 62\(\mu\)m \\ \hline
 \textbf{Wire Thickness} & 0.5\(\mu\)m & 0.5\(\mu\)m \\ \hline
 \textbf{Wire Pitch} & 30\(\mu\)m & 90\(\mu\)m \\ \hline
 \textbf{Wire Length} & 30mm (Cu) + & 30mm (Cu) + \\ 
  & 30mm (NbTiN) & 30mm (NbTiN) \\ \hline\hline
 \textbf{Data Rate} & 1Gbps & 1Gbps \\ \hline
 \textbf{No. of wires} & 20,000 & 10,000 \\ \hline
 \textbf{Required ML} & 2 & 8 \\ \toprule
\end{tabular}
}
\end{subfigure}
\caption{(\subref{fig:datalink}) Diagrammatic representation of the datalink interface (Cu over Glass bridge) connecting the 4K (Compute) and 77K (Main Memory) domains.
(\subref{table:datalink_spec})~Baseline specifications for the \mbox{main memory datalink}.}
\end{figure}

\underline{\textbf{Main Memory Datalink}}: A custom interface, as shown in Fig.~\ref{fig:datalink} is designed for data transfer between the 4K (Superconducting Compute) and 77K (Cryo-Memory) domains in our system. Physically, this dual-temperature interface connects the interposers in both domains via Cu transmission lines across a glass bridge. This datalink is DC coupled and thus requires no specific encoding. It performs translation for voltage levels, data rates, data direction and also protocol based on the amplification and PHY specifications at either ends(Fig.~\ref{table:datalink_spec}). For the given baseline specifications, our datalink can achieve a peak bidirectional bandwidth of 30TBps (20TBps Downlink and 10TBps Uplink). This can be increased or decreased based on the power budget, available metal layers, channel reach, reliability, noise \& dispersion etc.

\underline{\textbf{Cryo-DRAM block}}: The cryo-DRAM block consists of an array of cryo-DRAM packages integrated on a Si interposer. These have cryo-DRAM packages have no customizations or internal design changes and are simply regular DDR-X/LPDDR-X packages operating at 77K. As such these have inherent power benefits that have been extensively reported in~\cite{cryodram1,cryodram3,cryodram5}.

 A more detailed look into the power breakdown, micro-architecture and ISA lie outside the scope of this paper and will be pursued as future work.


\section{\textbf{Proposed System Architecture}}
\begin{figure*}[t]
    \centering
\includegraphics[width=0.93\linewidth]{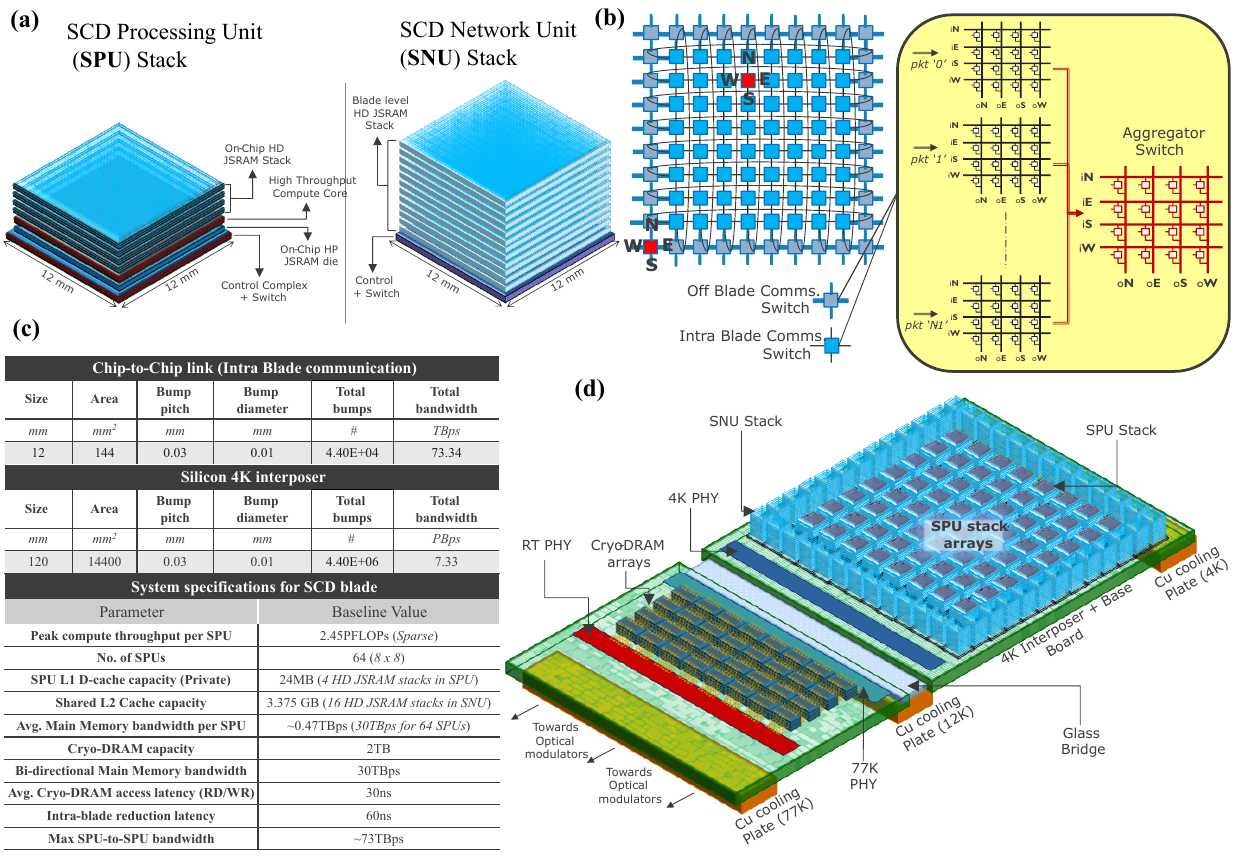}
    \caption{(a) Physical representation of the SPU and SNU stacks (b) Network topology and switch design (c) Baseline SCD Blade specifications for explorations (d) Physical representation of the full SCD blade.}
    \label{fig:blade_overview}
\end{figure*}
\subsection{SCD Processing and Network Unit (SPU and SNU)}
\noindent A single SPU (detailed in Fig.~\ref{fig:blade_overview}a), consists of a high compute throughput die (section~\ref{sec:designspec}), a host controller die, multiple HD JSRAM based Memory dies and HP JSRAM die all of which are vertically stacked by means of \texttt{NbTiN} through-silicon vias~(TSVs). The HD JSRAM dies serve both, the private L1 dcaches of the high throughput and control complex cores, as well as the shared L2 of the control complex. The HP JSRAM die contains the register files and L1 icaches for both the high throughput core and  control cores. The control complex as well as the local switch lie at the base of the SPU physical stack of dies. The SNU~(Fig.~\ref{fig:blade_overview}a) is another vertical stack of dies with a base die serving as switch for off-node or main memory communications. The JSRAM dies in each SNU die-stack are composed of banked HD arrays and function as slices of the shared and distributed L2 cache for all the high throughput cores (per SPU) in the blade. These help in bridging the latency gap for off-blade communication.

\subsection{SCD blade and interconnect}
\noindent We propose a SCD blade~(Fig.~\ref{fig:blade_overview}d), where, a 2D array of SPUs are interconnected via their local switches to construct a 2D torus intra-node network. The SNU die-stack lies at the edges of the 2D array of SPUs, and can even enable potential vertical stacking of SCD blades. This is made possible by extending \texttt{NbTiN}~TSVs in the SNU to connect to the substrate of the neighbouring blade (and optimizing its aspect ratio). Fig.~\ref{fig:blade_overview}c shows target specifications of the \texttt{Si} interposer, and intra communication interfaces. The performance numbers reported for the network interfaces are obtained considering bump density 4\%, bump redundancy 40\% and bandwidth per wire at 30~Gbps (30GHz operating frequency). In the proposed system, off-chip CryoDRAM operating at \texttt{77K} is connected to the compute array operating at \texttt{4K} by means of the datalink detailed in section~\ref{sec:designspec} and Fig.~\ref{fig:datalink}.

\subsection{System Architecture Parameters}
\label{sec:sys_arc}
\noindent  To assess the potential of SCDs for LLM workloads, we instantiate the system with the baseline parameters specified in Fig.~\ref{fig:blade_overview}c. Our blade consists of $8\times8$ SPUs (maximum \(\sim\)100 per blade, limited by the interposer stitching). The Cryo-DRAM capacity per blade is 2 TB accounting for $8\times 8$ quad die packages of LPPDDRx/DDRx bonded on a interposer~(77K) similar to the \texttt{Si} interposer housing SPU arrays. Every SPU can access the shared DRAM space at a bandwidth of 0.47 TBps. The entire $8\times 8$ blade can sustain a cumulative bi-directional main memory bandwidth of $30$ TBps at an average read/write latency of 30 ns. Due to the unique nature of superconducting interconnects and packaging technology, we can scale both the effective DRAM and network BW as we scale the number of SPUs.
The above specifications serve as the baseline  for the explorations on LLM training and inference via our performance model.

\section{\textbf{Analytical Performance Modelling}}

\begin{figure}[ht]
 \vspace{-0.3cm}
 \centering
   \includegraphics[width=0.85\linewidth]{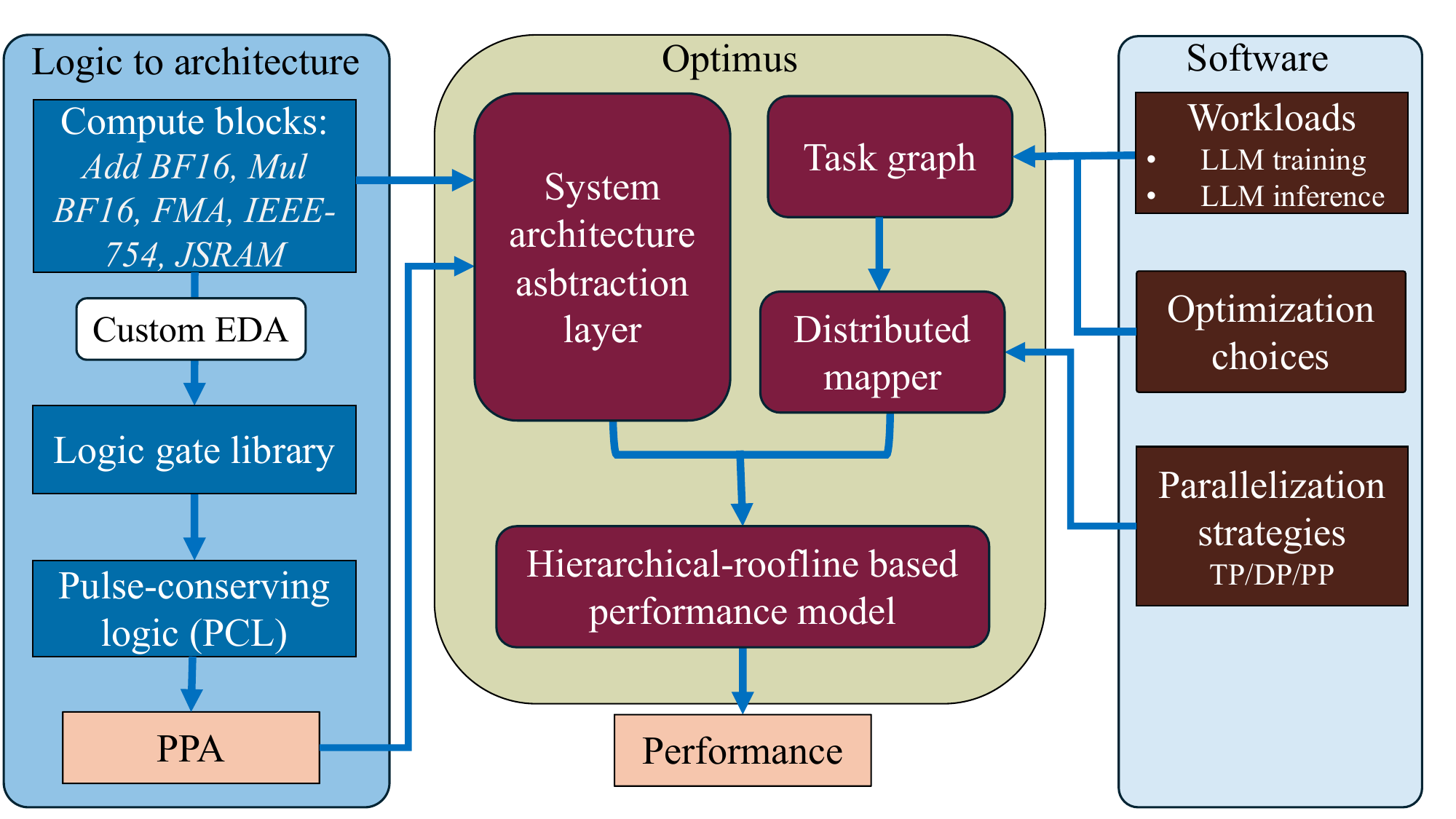}
    \caption{End-to-end performance analysis from logic design to system architecture, including mapping workload using task graphs for performance prediction for SCD system.}
     \label{fig:flow}
     \vspace{0.4cm}
 \end{figure}

\noindent In this section, we describe the principles of the performance modeling approach (Fig.~\ref{fig:flow}). The measured SCD technology data (section~\ref{sec:background}) is used to create the basic design blocks, which are further used to derive the high-level system architectural parameters, shown in Fig.~\ref{fig:blade_overview}c. 

Given a workload (e.g., LLM training/inference), the modeling framework, Optimus, ingests a detailed task graph with the LLM model parameters such as number of layers, attention heads, hidden dimension, input/output shapes, sequence length, batch-size, working precision, etc. Using the above parameters and a chosen combination of parallelization strategies, such as data parallelism~(DP), tensor model parallelism~(TP) and pipeline parallel~(PP), the workload is mapped onto the underlying system architecture.
In DP, the model is replicated and data is sharded; in TP the model is sharded and data is replicated and in PP the model is sharded layer wise and data is divided into small chunks to inject in a pipeline fashion. Each parallelism option comes with its own advantages and trade offs~\cite{AMPeD}.
At its core, Optimus relies on a hierarchical roofline model for a single accelerator to determine if a given kernel in the task graph is compute or memory (on-chip/off-chip) bound. For compute bound kernels, the execution time is primarily determined by the compute throughput, while for memory-bound kernels, it is dominated by the data transfer time from the respective memory level. For a given system architecture and workload, we assess the most optimal mapping, reducing communication overhead~\cite{eff_megatron}. The performance modeling framework has been validated against both training and inference on different GPUs, for different parallelization strategies over multiple LLM models~\cite{optimus_arxiv}. In this work, we adapt Optimus for SCD system architecture. 



\section{\textbf{Results and Discussion}}
\noindent Here, we present the projected performance numbers using the performance modeling framework. As described in the previous sections, the SCD system has a huge memory and network bandwidth advantage compared to contemporary systems, which memory bound applications can exploit. To test this hypotheses, we model both LLM training and inference on the SCD system and project its performance in comparison with equivalent number of GPUs~(H100s: peak throughput of 0.9895 PFLOPs, DRAM bandwidth of 3.35 TBps). 

\begin{figure}
    \centering
\includegraphics[width=0.9\linewidth]{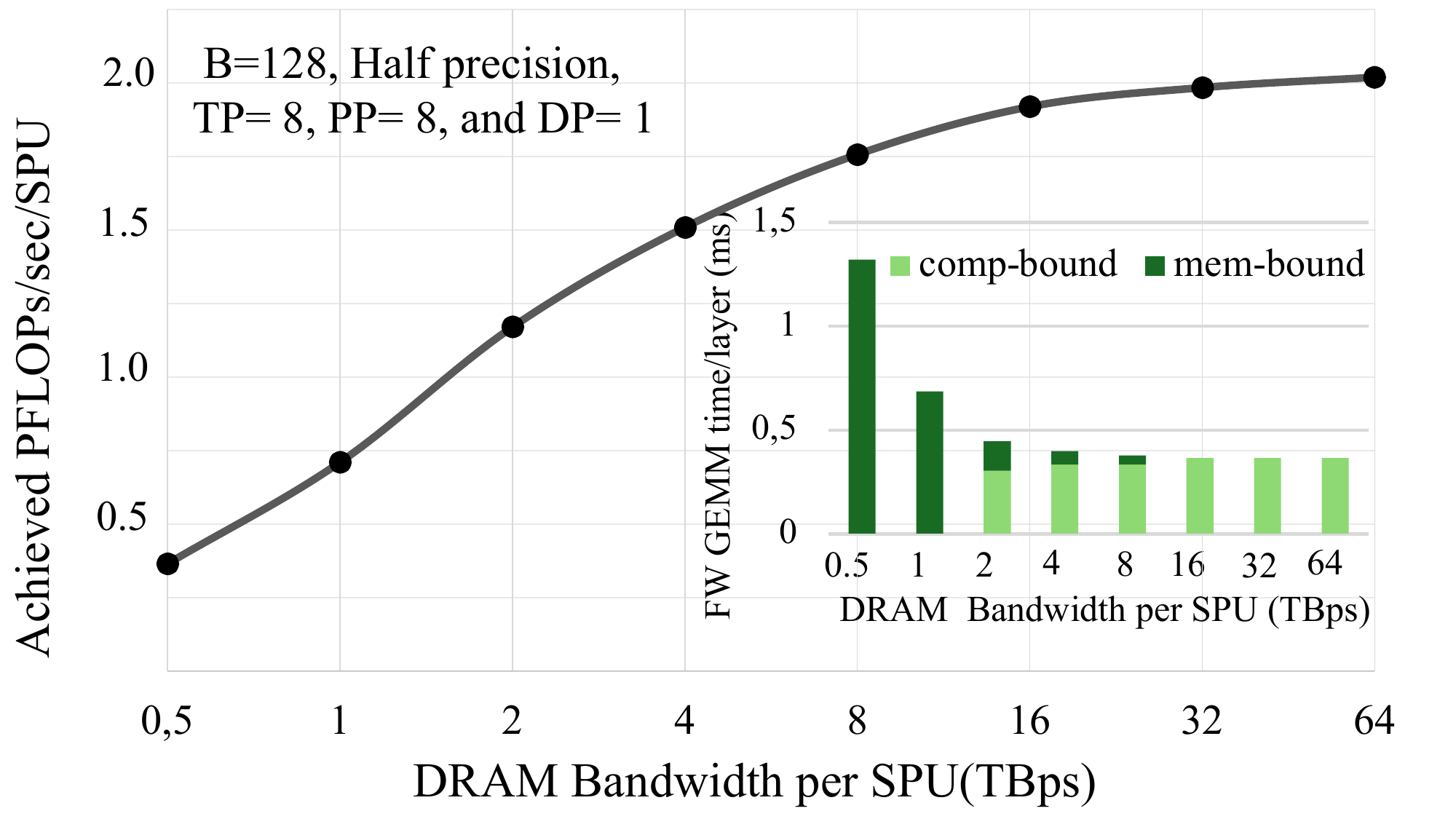}
    \caption{Impact of DRAM bandwidth~(BW) per SPU in SCD system on the achieved throughput (PFLOPs/SPU) per batch for GPT3-76B model training. Inset shows the time breakdown of the memory versus compute-bound GEMMs per layer in the forward pass.} 
    \label{fig:bw_scaing}
\end{figure}
First, we look at LLM training to investigate the potential speed-up of SCD systems. The majority of the matrix multiplication~(GEMM) kernels in LLM training are compute-bound on GPUs; however, the GEMMs involved in the attention computation and non-linear operations are typically memory bandwidth bound~\cite{flashatt2022} (depends on kernel flop-to-byte ratio and hardware compute throughput-to-bandwidth ratio). Fig.~\ref{fig:bw_scaing}, gives the achieved throughput in PFLOPs/SPU for training GPT3-(76B model) on 64 SPUs, considering fixed model parallelism with a TP degree of 8 and a PP degree of 8. We perform an exploration of the total effective bandwidth available to an SPU. The minimum memory bandwidth available to each SPU is 0.47~TBps as discussed in Sec.~\ref{sec:sys_arc}~(Fig.~\ref{fig:blade_overview}c). We then perform a sweep for the memory bandwidth available per SPU from0.5~TBps to 64~TBps and estimate the achieved throughput. As seen in Fig.~\ref{fig:bw_scaing}, the performance scales with bandwidth monotonically and tend to saturate around 2~PFLOPs/SPU at 16~TBps of bandwidth. To further analyze this, we look at the compute versus memory-boundedness of GEMMs involved in the forward pass as a function of the memory bandwidth. As seen in the inset, at lower bandwidths, most of the GEMM time is spent in memory-bound kernels,and thus performance can improve when bandwidth is increased. We observe a gradual crossover from a memory-bound scenario to compute-bound one for a memory bandwidth \(\geq\)16~TBps per SPU. The modest improvement in the performance beyond 16~TBps is due to the remaining memory-bound operations in the task-graph, for instance, non-linear operations like, softmax, layer-norm etc. 
\begin{figure}
    \centering
\includegraphics[width=0.9\linewidth]{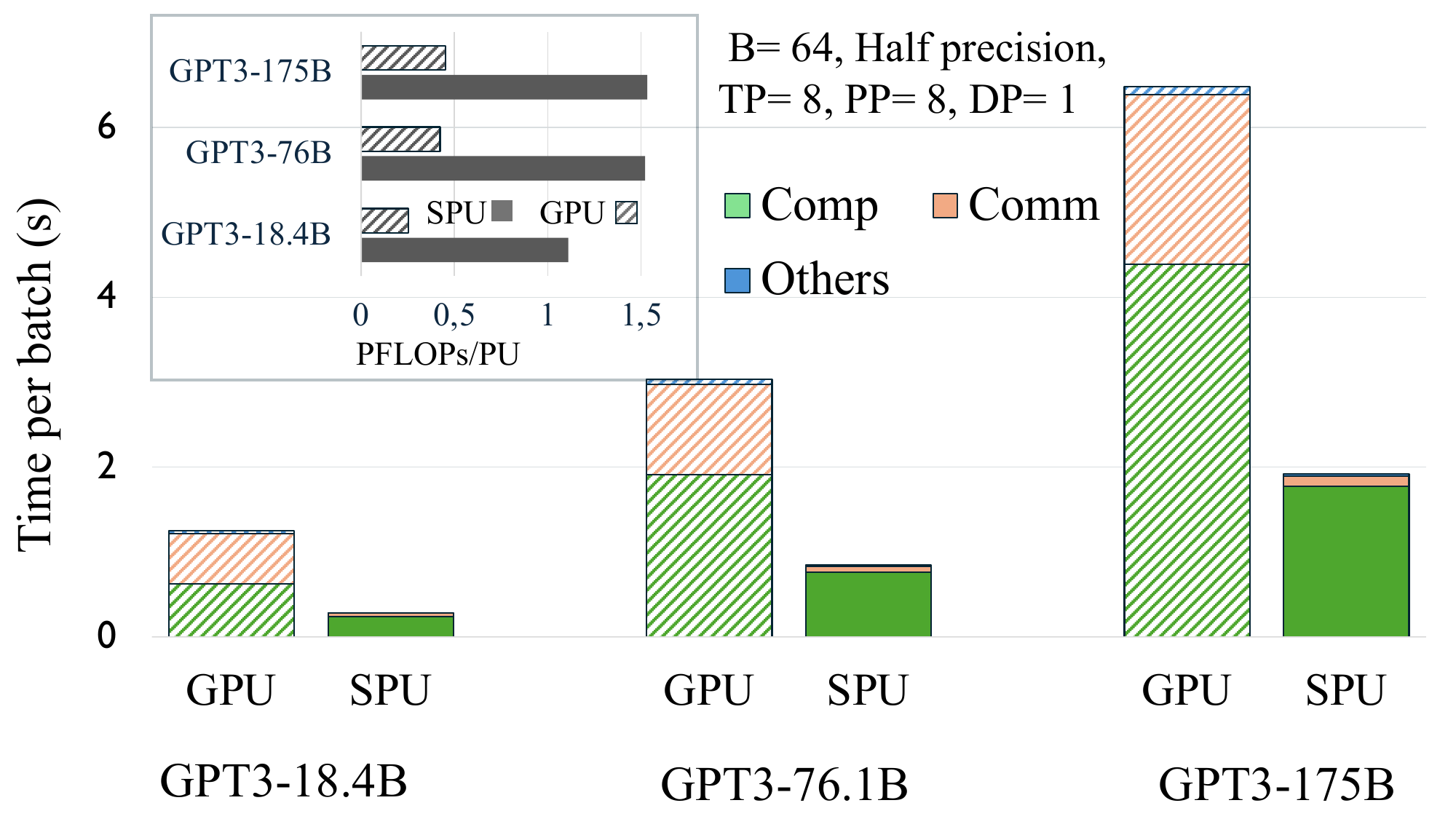}
    \caption{Processing time  per batch with a break up of compute, communication, and remaining time for three different GPT models in training for GPU (H100) (patterned) and SPU (solid). The inset shows the corresponding throughput in PFLOPs/processing unit.} 
    \label{fig:training}
    \vspace{0.5cm}
\end{figure}

\begin{figure}[t]
    \centering
    \includegraphics[width=0.86\linewidth]{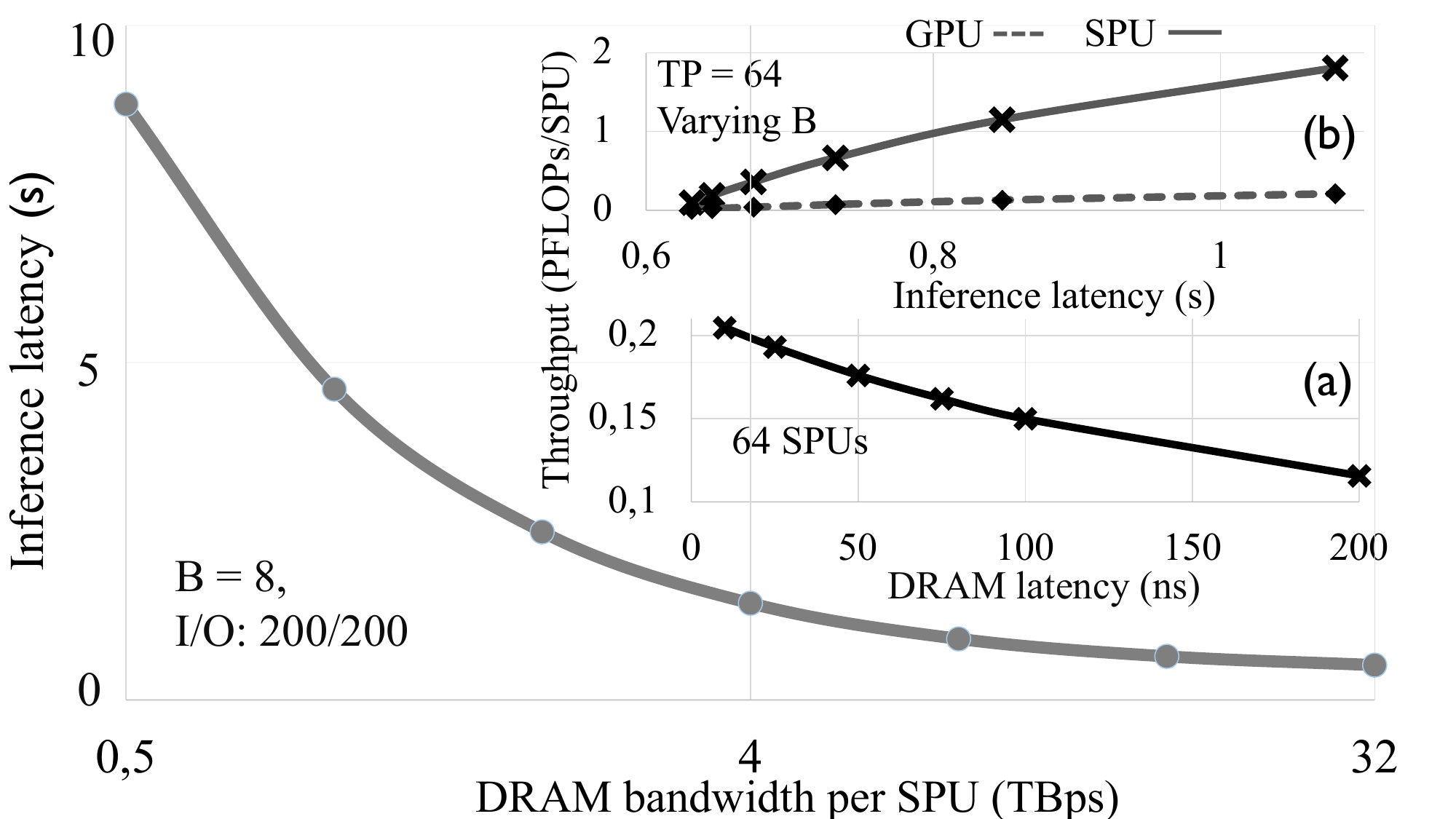}
    \caption{Impact of DRAM bandwidth on inference latency(s) for Llama-405B model. We set bf16 precision and I/O: 200/200 tokens, DRAM latency=30ns. Inset (a): impact of DRAM latency on the achieved throughput (B = 8). Inset (b): inference latency versus achieved throughput while B is varied from 4 to 128. For the insets, the effective DRAM bandwidth per SPU is set at 16 TBps.}
    \label{fig:single-blade}
\end{figure}
\begin{figure}[t]
    \centering
    \includegraphics[width=0.9\linewidth]{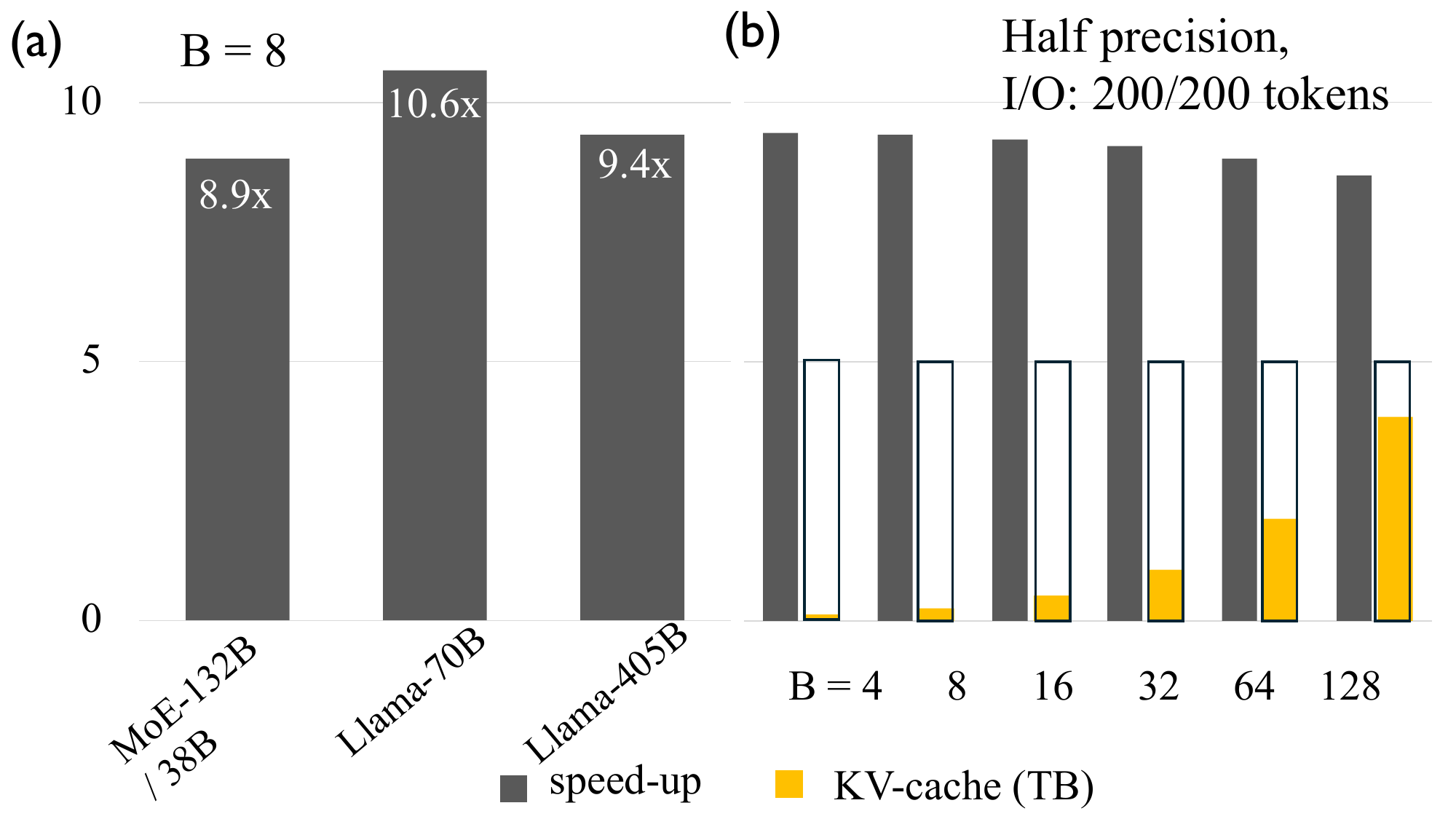}
    \caption{(a) Single SCD blade (with 64 SPUs) inference speed-up  (vs the same number of GPUs) for three different LLM models. (b) Impact of batch size on the speed-up and the respective KV-cache size for Llamma-405B model. Height of the open bar denote the total memory capacity of 64 GPUs for reference (=64$\times$80 GB). Memory bandwidth per SPU is set to 16TBps.}
    \label{fig:inference_models}
\end{figure}
\noindent \underline{\textbf{Key takeaway}}: {\em Achieved throughput grows with available bandwidth initially and slowly saturates as most kernels turn compute-bound vs being memory-bound.}

We now compare the performance of training different GPT models on SPUs vs the same number of GPUs.Fig.~\ref{fig:training}, shows the break up of compute, communication and others (pipeline bubble + weight update time) for three different models (for TP=8, PP=8, DP=1 and B=64 and bf16 precision). We set the total effective bandwidth available to each SPU in a blade (64 SPUs) at 16 TBps. In all the three cases, we observe that the SCD system is significantly faster than the GPU system (64 H100s). The speed-up varies from $3.5\times$ -- $4.4\times$ for this particular set up and model sizes. As expected, the SCD system enables both faster compute and faster communication -- the primary gain coming from faster data movement. The inset exhibits the achieved throughput in terms of PFLOPs per processing units for GPU and SCD systems with the same experimental set up. We should note that in this case the maximal achieved throughput is around 1.5~PFLOPs/SPU while in the previous set up it is around 2~PFLOPs/SPU. This is primarily due the increased batch size (64 versus 128), that leads to improved throughput for both GPU and SPU. However, the SCD systems benefit more where the data transfer overhead is larger. 

\noindent \underline{\textbf{Key takeaway}}: {\em SCD system promises to offer significantly higher throughput compared to contemporary GPU systems across different LLM models for training due to higher cumulative memory and interconnect bandwidth. }

We now move to LLM inference. Inference is known to be a memory-bound workload due to minimal data reuse in the involved kernels~\cite{optimus_arxiv} and thus, we expect SCD systems to perform better in this case vs LLM training. In, the following study, we vary the effective DRAM bandwidth per SPU from 0.5 TBps to 32 TBps to see its impact on the inference latency.
Fig.~\ref{fig:single-blade} shows how the inference latency goes down as we increase the bandwidth seen by each SPU. Going from 0.5 TBps~(8.8s) to 32 TBps ~(0.52s), we see a speed up of $17\times$, thus confirming the importance of memory bandwidth in inference. The performance scaling tend to saturate beyond 8 TBps since we start hitting the DRAM latency bound limit. Here, we consider the Llama-405B model with a batch size of 8, bf16 precision, summarization and prediction of 200 tokens; DRAM latency is set to 30ns. Tensor parallelism is utilized here, where the number of SPUs is same as the TP degree. 

Since the collective DRAM bandwidth to a blade can be very high for the given SCD architecture, we analyze the sensitivity of the achieved throughput with respect to DRAM latency keeping the effective DRAM bandwidth for each SPU at 16 TBps. Inset (a) in Fig.~\ref{fig:single-blade} shows that the achieved PFLOPs/SPU goes down almost linearly with increasing DRAM latency (varied from 10ns-- 200ns). We see that lower DRAM latency is crucial to achieve high performance in the case of inference. Inset (b) shows the variation of achieved throughput with inference latency (H100 versus SPU) as we vary the batch size from 4 to 128 (DRAM latency is set at 30ns). As the batch size is increased, the achieved PFLOPs/SPU increases along with the inference latency-- this trade off helps determining the number of queries that can be batched without sacrificing user experience.  

\noindent \textbf{\underline{Key takeaway}}: {\em LLM inference performance continue to scale with increase in available DRAM bandwidth when the DRAM latency is low.}

Next, we investigate the speed-up in inference across different LLM models (MoE-132B/38B, Llama-70B, Llama-405B) for different batch sizes at a single SCD blade level compared to equivalent number of GPUs (H100s). We see a massive speed-up of 9$\times$-11$\times$ depending on the LLM model (see Fig.~\ref{fig:inference_models}a, DRAM latency is set to 30 ns and bandwidth to 16~TBps). SCD performs best for Llama-70B among these models since the fraction of communication overhead is maximal in this case. For the Mixture-of-Experts model, the speed-up is a bit lower than that for Llama models since only 4 experts are active among 16. Hence, the communication overhead is less compared to the other models. The right panel shows the robustness of the speed-up with respect to the batch size for Llama-405B model. In addition, we also show the required KV-cache size as a function of batch size. We see that, at a batch size of 128, the KV-cache size is very close the maximum memory capacity of 64 GPUs (5TB), thus potentially limiting scaling up of batch sizes further for GPU systems. The speedup of our SCD system gradually decreases for large batch sizes (till we hit the memory capacity limit of GPUs) relative to GPU systems since the compute to data-transfer ratio now goes up.

\noindent \textbf{\underline{Key takeaway}}: {\em The SCD system offers even more performant execution of LLM inference compared to training on GPU systems (due to memory boundedness of inference). The speed-up is robust across different LLM models and batch sizes.}

Lastly, we explore the potential use of the large L2 caches ($\sim 4.19$ GB). The required kv-cache size for the popular llama models are, llama2-7B: 2 GB, llama2-13B: 3 GB and llama2-70B: 10 GB. 
Thus, one can possibly fit the entire kv-cache of the two smaller llama models (7B and 13B) onto the L2 cache of the SCD architecture. For GPU systems, the cache size is in the order of MBs (e.g., 50 MB for H100) and cannot be shared across GPUs. Thus, the GEMMs/GEMVs involving {\em Key} and {\em Value} can be accelerated by making them L2 or compute bound from a traditional DRAM bound case. Since L2 cache also comes with a huge bandwidth jump compared to DRAM, one can exploit it by properly managing the memory. Our early estimates suggest a speed-up of $\sim2-4\times$ for the relevant GEMMs/GEMVs (depending on the software overhead of launching the kernels) if such a scheme is implemented. Since we do not have a clear estimate of the software overhead in those cases, it is hard to model the impact accurately.

\section{\textbf{Conclusions and future outlook}}
In this paper, we present a comprehensive study on performance modeling of SCD technology for LLMs training and inference using a bottom-up approach to derive high level system architectural parameters based superconducting process and device level data. We show that the significantly high memory and interconnect bandwidth of the SCD technology can substantially accelerate both LLM training and inference, serving as a very promising solution to the memory and interconnect wall problems. Although, we limit this study to projecting the performance of a single SCD blade, we expect the performance to scale with the number of blades -- to be explored in our future investigations. Another interesting direction could be to look at the impact of huge JSRAM capacity on LLM inference exploiting its massive bandwidth and negligible latency. Such unusual SRAM capacity will further lead to new ways of mapping and memory management approaches. 




\bibliography{references}

\begin{thebibliography}{10}
\providecommand{\url}[1]{#1}
\csname url@samestyle\endcsname
\providecommand{\newblock}{\relax}
\providecommand{\bibinfo}[2]{#2}
\providecommand{\BIBentrySTDinterwordspacing}{\spaceskip=0pt\relax}
\providecommand{\BIBentryALTinterwordstretchfactor}{4}
\providecommand{\BIBentryALTinterwordspacing}{\spaceskip=\fontdimen2\font plus
\BIBentryALTinterwordstretchfactor\fontdimen3\font minus
  \fontdimen4\font\relax}
\providecommand{\BIBforeignlanguage}[2]{{%
\expandafter\ifx\csname l@#1\endcsname\relax
\typeout{** WARNING: IEEEtran.bst: No hyphenation pattern has been}%
\typeout{** loaded for the language `#1'. Using the pattern for}%
\typeout{** the default language instead.}%
\else
\language=\csname l@#1\endcsname
\fi
#2}}
\providecommand{\BIBdecl}{\relax}
\BIBdecl

\bibitem{McKinsey_LLM}
M.~. Company, ``{What’s the future of generative AI? An early view in 15
  charts},''
  \url{https://www.mckinsey.com/featured-insights/mckinsey-explainers/whats-the-future-of-generative-ai-an-early-view-in-15-charts},
  2023.

\bibitem{HPCwire_LLM}
\BIBentryALTinterwordspacing
A.~Solon. (2023) Generative ai boosts demand for compute resources: Altman
  solon analysis. [Online]. Available:
  \url{https://www.hpcwire.com/off-the-wire/generative-ai-boosts-demand-for-compute-resources-altman-solon-analysis/}
\BIBentrySTDinterwordspacing

\bibitem{transformer_paper}
A.~Vaswani, N.~Shazeer, N.~Parmar, J.~Uszkoreit, L.~Jones, A.~N. Gomez,
  {\L}.~Kaiser, and I.~Polosukhin, ``Attention is all you need,'' in
  \emph{Advances in Neural Information Processing Systems}, 2017, pp.
  5998--6008.

\bibitem{NIST_Energy}
J.~Booth, ``\BIBforeignlanguage{en}{Energy efficiency scaling for 2 decades
  (ees2) roadmap for computing},'' in \emph{\BIBforeignlanguage{en}{28th Annual
  IEEE High Performance Extreme Computing Virtual Conference, Virtual, CO,
  US}}.\hskip 1em plus 0.5em minus 0.4em\relax IEEE, 2024-07-14 00:07:00 2024.

\bibitem{paterson2022}
D.~Patterson, J.~Gonzalez, U.~Hölzle, Q.~Le, C.~Liang, L.-M. Munguia,
  D.~Rothchild, D.~R. So, M.~Texier, and J.~Dean, ``The carbon footprint of
  machine learning training will plateau, then shrink,'' \emph{Computer},
  vol.~55, no.~7, pp. 18--28, 2022.

\bibitem{paterson2021}
\BIBentryALTinterwordspacing
D.~A. Patterson, J.~Gonzalez, Q.~V. Le, C.~Liang, L.~Munguia, D.~Rothchild,
  D.~R. So, M.~Texier, and J.~Dean, ``Carbon emissions and large neural network
  training,'' \emph{CoRR}, vol. abs/2104.10350, 2021. [Online]. Available:
  \url{https://arxiv.org/abs/2104.10350}
\BIBentrySTDinterwordspacing

\bibitem{Inference_cost}
S.~Samsi, D.~Zhao, J.~McDonald, B.~Li, A.~Michaleas, M.~Jones, W.~Bergeron,
  J.~Kepner, D.~Tiwari, and V.~Gadepally, ``From words to watts: Benchmarking
  the energy costs of large language model inference,'' in \emph{2023 IEEE High
  Performance Extreme Computing Conference (HPEC)}, 2023, pp. 1--9.

\bibitem{AI_Memory}
A.~Gholami, Z.~Yao, S.~Kim, C.~Hooper, M.~W. Mahoney, and K.~Keutzer, ``{AI and
  Memory Wall},'' \emph{IEEE Micro}, vol.~44, no.~3, pp. 33--39, 2024.

\bibitem{chip_TCO}
\BIBentryALTinterwordspacing
N.~Asia. (2023) The great nanometer chip race. [Online]. Available:
  \url{https://asia.nikkei.com/Spotlight/The-Big-Story/The-great-nanometer-chip-race}
\BIBentrySTDinterwordspacing

\bibitem{Memory_wall}
\BIBentryALTinterwordspacing
S.~A. McKee, ``Reflections on the memory wall,'' in \emph{Proceedings of the
  1st Conference on Computing Frontiers}, ser. CF '04.\hskip 1em plus 0.5em
  minus 0.4em\relax ACM, 2004, p. 162. [Online]. Available:
  \url{https://doi.org/10.1145/977091.977115}
\BIBentrySTDinterwordspacing

\bibitem{Interconnect_wall}
K.~Banerjee and A.~Mehrotra, ``Global (interconnect) warming,'' \emph{IEEE
  Circuits and Devices Magazine}, vol.~17, no.~5, pp. 16--32, 2001.

\bibitem{Scaling_laws}
\BIBentryALTinterwordspacing
J.~Kaplan, S.~McCandlish, T.~Henighan, T.~B. Brown, B.~Chess, R.~Child,
  S.~Gray, A.~Radford, J.~Wu, and D.~Amodei, ``Scaling laws for neural language
  models,'' \emph{CoRR}, vol. abs/2001.08361, 2020. [Online]. Available:
  \url{https://arxiv.org/abs/2001.08361}
\BIBentrySTDinterwordspacing

\bibitem{29_pcl_imec_asc_22}
T.~Josephsen and Q.~Herr, ``Pulse conserving logic,'' \emph{poster presented at
  Applied Superconducting Conference}, 2022.

\bibitem{27_sc_processor_2018}
M.~Vesely, P.~Tschirhart, B.~Konigsburg, P.~Farrell, S.~Rahman, and E.~Massaad,
  ``A pipelined superconducting 16-bit cpu design,'' \emph{Applied
  Superconducting Conference}, 2018.

\bibitem{30_scinterposer_2022}
\BIBentryALTinterwordspacing
J.~Egan, M.~Nielsen, J.~Strong, V.~Talanov, E.~Rudman, B.~Song, Q.~Herr, and
  A.~Herr, ``Synchronous chip-to-chip communication with a multi-chip resonator
  clock distribution network*,'' \emph{Superconductor Science and Technology},
  vol.~35, no.~10, p. 105010, sep 2022. [Online]. Available:
  \url{https://dx.doi.org/10.1088/1361-6668/ac8e38}
\BIBentrySTDinterwordspacing

\bibitem{bairamkulov2024superconductive}
R.~Bairamkulov and G.~De~Micheli, ``Superconductive electronics: A 25-year
  review [feature],'' \emph{IEEE Circuits and Systems Magazine}, vol.~24,
  no.~2, pp. 16--33, 2024.

\bibitem{ayala2020mana}
C.~L. Ayala, T.~Tanaka, R.~Saito, M.~Nozoe, N.~Takeuchi, and N.~Yoshikawa,
  ``Mana: A monolithic adiabatic integration architecture microprocessor using
  1.4-zj/op unshunted superconductor josephson junction devices,'' \emph{IEEE
  Journal of Solid-State Circuits}, vol.~56, no.~4, pp. 1152--1165, 2020.

\bibitem{herr2023superconducting}
Q.~Herr, T.~Josephsen, and A.~Herr, ``Superconducting pulse conserving logic
  and josephson-sram,'' \emph{Applied Physics Letters}, vol. 122, no.~18, 2023.

\bibitem{egan2022synchronous}
J.~Egan, M.~Nielsen, J.~Strong, V.~Talanov, E.~Rudman, B.~Song, Q.~Herr, and
  A.~Herr, ``Synchronous chip-to-chip communication with a multi-chip resonator
  clock distribution network,'' \emph{Superconductor Science and Technology},
  vol.~35, no.~10, p. 105010, 2022.

\bibitem{dai2022isochronous}
H.~Dai, C.~Kegerreis, D.~W. Gamage, J.~Egan, M.~Nielsen, Y.~Chen, D.~Tuckerman,
  S.~E. Peek, B.~Yelamanchili, M.~Hamilton \emph{et~al.}, ``Isochronous data
  link across a superconducting nb flex cable with 5 femtojoules per bit,''
  \emph{Superconductor Science and Technology}, vol.~35, no.~4, p. 045014,
  2022.

\bibitem{herr2024data}
A.~Herr and Q.~Herr, ``A data center in a shoebox: Imec's plan to use
  superconductors to shrink computers,'' \emph{IEEE Spectrum}, vol.~61, no.~6,
  pp. 37--41, 2024.

\bibitem{Nb_JJ1}
S.~K. Tolpygo, V.~Bolkhovsky, S.~Zarr, T.~J. Weir, A.~Wynn, A.~L. Day, L.~M.
  Johnson, and M.~A. Gouker, ``Properties of unshunted and resistively shunted
  nb/alox-al/nb josephson junctions with critical current densities from 0.1 to
  1 ma/$\mu$m2,'' \emph{IEEE Transactions on Applied Superconductivity},
  vol.~27, no.~4, pp. 1--15, 2017.

\bibitem{Nb_JJ2}
S.~K. Tolpygo, V.~Bolkhovsky, R.~Rastogi, S.~Zarr, A.~L. Day, E.~Golden, T.~J.
  Weir, A.~Wynn, and L.~M. Johnson, ``Advanced fabrication processes for
  superconductor electronics: Current status and new developments,'' \emph{IEEE
  Transactions on Applied Superconductivity}, vol.~29, no.~5, pp. 1--13, 2019.

\bibitem{Nb_JJ3}
J.~K. Sergey~Tolpygo, Thomas~Schurig, ``Processing and manufacture of josephson
  junctions: Low-t c,'' in \emph{Handbook of Superconductivity}.\hskip 1em plus
  0.5em minus 0.4em\relax CRC Press, 2021.

\bibitem{JJscaling1}
\BIBentryALTinterwordspacing
S.~Steinhauer, L.~Yang, S.~Gyger, T.~Lettner, C.~Errando-Herranz, K.~D. Jöns,
  M.~A. Baghban, K.~Gallo, J.~Zichi, and V.~Zwiller, ``{NbTiN thin films for
  superconducting photon detectors on photonic and two-dimensional
  materials},'' \emph{Applied Physics Letters}, vol. 116, no.~17, p. 171101, 04
  2020. [Online]. Available: \url{https://doi.org/10.1063/1.5143986}
\BIBentrySTDinterwordspacing

\bibitem{JJscaling2}
\BIBentryALTinterwordspacing
L.~Zhang, W.~Peng, L.~X. You, and Z.~Wang, ``{Superconducting properties and
  chemical composition of NbTiN thin films with different thickness},''
  \emph{Applied Physics Letters}, vol. 107, no.~12, p. 122603, 09 2015.
  [Online]. Available: \url{https://doi.org/10.1063/1.4931943}
\BIBentrySTDinterwordspacing

\bibitem{SCDprocess}
V.~Vega-Gonzalez, D.~Radisic, S.~Choudhury, D.~Tierno, A.~Thiam, D.~Batuk,
  G.~Martinez, F.~Seidel, S.~Decoster, S.~Kundu, D.~Tsvetanova, A.~Peter,
  H.~De~Coster, A.~Sepulveda-Marquez, E.~Altamirano-Sanchez, B.~Chan,
  Y.~Drissi, Y.~Sherazi, J.~Uk-Lee, I.~Ciofi, G.~Murdoch, N.~Nagesh,
  G.~Hellings, J.~Ryckaert, S.~Biesemans, E.~D. Litta, N.~Horiguchi, S.~Park,
  and Z.~Tökei, ``Semi-damascene integration of a 2-layer mol vhv scaling
  booster to enable 4-track standard cells,'' in \emph{2022 International
  Electron Devices Meeting (IEDM)}, 2022, pp. 23.2.1--23.2.4.

\bibitem{SCDprocess2}
G.~Murdoch, Z.~Tokei, S.~Paolillo, O.~V. Pedreira, K.~Vanstreels, and C.~J.
  Wilson, ``Semidamascene interconnects for 2nm node and beyond,'' in
  \emph{2020 IEEE International Interconnect Technology Conference (IITC)},
  2020, pp. 4--6.

\bibitem{herr2023scaling}
A.~Herr, Q.~Herr, S.~Brebels, M.-S. Kim, A.~Pokhrel, B.~Hodges, T.~Josephsen,
  S.~ONeal, R.~Bai, K.~Nowack \emph{et~al.}, ``Scaling nbtin-based ac-powered
  josephson digital to 400m devices/cm $^2$,'' \emph{arXiv preprint
  arXiv:2303.16792}, 2023.

\bibitem{cryodram1}
F.~Wang, T.~Vogelsang, B.~Haukness, and S.~C. Magee, ``Dram retention at
  cryogenic temperatures,'' in \emph{2018 IEEE International Memory Workshop
  (IMW)}, 2018, pp. 1--4.

\bibitem{cryodram3}
G.-H. Lee, D.~Min, I.~Byun, and J.~Kim, ``Cryogenic computer architecture
  modeling with memory-side case studies,'' in \emph{2019 ACM/IEEE 46th Annual
  International Symposium on Computer Architecture (ISCA)}, 2019, pp. 774--787.

\bibitem{cryodram5}
\BIBentryALTinterwordspacing
F.~Ware, L.~Gopalakrishnan, E.~Linstadt, S.~A. McKee, T.~Vogelsang, K.~L.
  Wright, C.~Hampel, and G.~Bronner, ``Do superconducting processors really
  need cryogenic memories? the case for cold dram,'' in \emph{Proceedings of
  the International Symposium on Memory Systems}, ser. MEMSYS '17.\hskip 1em
  plus 0.5em minus 0.4em\relax New York, NY, USA: Association for Computing
  Machinery, 2017, p. 183–188. [Online]. Available:
  \url{https://doi.org/10.1145/3132402.3132424}
\BIBentrySTDinterwordspacing

\bibitem{AMPeD}
D.~Moolchandani, J.~Kundu, F.~Ruelens, P.~Vrancx, T.~Evenblij, and
  M.~Perumkunnil, ``Amped: An analytical model for performance in distributed
  training of transformers,'' in \emph{2023 IEEE International Symposium on
  Performance Analysis of Systems and Software (ISPASS)}.\hskip 1em plus 0.5em
  minus 0.4em\relax IEEE, 2023, pp. 306--315.

\bibitem{eff_megatron}
D.~Narayanan, M.~Shoeybi, J.~Casper, P.~LeGresley, M.~Patwary, V.~Korthikanti,
  D.~Vainbrand, P.~Kashinkunti, J.~Bernauer, B.~Catanzaro \emph{et~al.},
  ``Efficient large-scale language model training on gpu clusters using
  megatron-lm,'' in \emph{Proceedings of the International Conference for High
  Performance Computing, Networking, Storage and Analysis}, 2021, pp. 1--15.

\bibitem{optimus_arxiv}
\BIBentryALTinterwordspacing
J.~Kundu, W.~Guo, A.~BanaGozar, U.~D. Alwis, S.~Sengupta, P.~Gupta, and
  A.~Mallik, ``Performance modeling and workload analysis of distributed large
  language model training and inference,'' in \emph{2024 IEEE International
  Symposium on Workload Characterization (IISWC)}, 2024. [Online]. Available:
  \url{https://arxiv.org/abs/2407.14645}
\BIBentrySTDinterwordspacing

\bibitem{flashatt2022}
\BIBentryALTinterwordspacing
T.~Dao, D.~Y. Fu, S.~Ermon, A.~Rudra, and C.~Ré, ``Flashattention: Fast and
  memory-efficient exact attention with io-awareness,'' 2022. [Online].
  Available: \url{https://arxiv.org/abs/2205.14135}
\BIBentrySTDinterwordspacing

\end{thebibliography}
\end{document}